\documentclass[
 reprint,
superscriptaddress,
 amsmath,amssymb,
 aps,
prb
]{revtex4-2}

\usepackage{pifont}
\usepackage{graphicx}
\usepackage{bm}
\usepackage{sidecap}
\usepackage{xcolor}
\usepackage[colorlinks=true,citecolor=blue]{hyperref}
\usepackage{natbib}

\makeatletter

\newcommand{\Rmnum}[1]{\expandafter\@slowromancap\romannumeral #1@}
\makeatother

\begin{document}

\title{Janus bound states in the continuum in structurally symmetric photonic crystals}

\author{Hongzhi Zuo}
\affiliation{School of Physics and Optoelctronics, Xiangtan University, Xiangtan 411105, China}

\author{Shengxuan Xia}
\affiliation{School of Physics and elctronics, Hunan University, Changsha 410082, China}
\affiliation{State Key Laboratory of Millimeter Waves, Southeast University, Nanjing 210096, China}

\author{Haiyu Meng}
\email{H.Y.Meng@hnu.edu.cn}
\affiliation{School of Physics and Optoelctronics, Xiangtan University, Xiangtan 411105, China}

\begin{abstract}
We propose a $\sigma_z$-symmetry-preserving approach to achieve Janus bound states in the continuum (Janus BICs) exhibiting asymmetric topological charges in the upward and downward radiation channels. While prior approaches typically involve explicit structural perturbations to break vertical symmetry, our design leverages a bilayer photonic crystal slab (PCS) system with independently tunable refractive indices, introducing an optical asymmetry without altering the geometric symmetry. In the optical symmetry case, the system supports symmetry-protected BICs at $\Gamma$ point with topological charge $q = -1$, and Friedrich–Wintgen BICs (FW-BICs) at off-$\Gamma$ point with $q = +1$. Upon introducing refractive index detuning, the polarization vortex splits into two circularly polarized states (C points) with half-integer topological charge ($q = 1/2$), shifting oppositely in momentum space for upward and downward radiation, while the symmetry-protected BICs remain unaffected. Janus BICs are established through the shift of upward-radiating C points shift towards the $\Gamma$ point, accumulating a total topological charge of $q = -1$, while the downward-radiating counterpart contributes $q = +1$, leading to a net topological charge reversal between the two radiation channels. This purely optical mechanism allows for the realization of Janus BICs without any structural deformation. Their asymmetric topological nature makes them ideally suited for applications in unidirectional light sources, chiral photonic interfaces, and topological photonic circuits, offering a promising platform for on-chip optical communication, sensing, and quantum information processing. 
\end{abstract}
\maketitle

Topological defects, such as polarization singularities and phase vortices, have become central topics in modern photonics, owing to their robustness, topological protection, and potential applications in light manipulation, sensing, and information processing  \cite{zhen2014topological,kang2025janus,chen2019observing,lee2014lattice}. Among these, polarization vortices play a pivotal role by shaping the optical field topology in momentum space, characterized by nonzero topological charges. In photonic systems, including photonic crystal slab (PCS) \cite{lin2017line,zhou2024dual,zhang2019entangled,ni2024three} and metamaterials \cite{qian2024manipulation,zhong2024toroidal,meng2022terahertz,dai2024quasi,tan2021active,meng2023port}, these vortices not only shed light on the intricate relationship between topology and light–matter interactions but also offer new possibilities for advanced optical control. A particularly compelling manifestation of topological features in photonics is the realization of bound states in the continuum (BICs), which enable the confinement of light within open structures without radiation loss \cite{zhou2023quasi,marinica2008bound,kang2023applications,tittl2018imaging,hsu2016bound,kang2022merging,hu2022global,jiang2023general,kang2022coherent,zhang2025merging,zhou2025bound,zhang2025multiple,zhang2025perturbation}. BICs are non-radiative modes embedded in the continuous spectrum of leaky states, supporting theoretically infinite quality (Q) factors  \cite{yi2025efficient,barkaoui2023merged,cui2023single,gao2024degenerate,sun2024tailoring,liu2024merging,wu2024momentum,li2025ultrasensitive,meng2022bound,cao2023interaction,qian2024non,liu2023boosting,wan2025photonic}. They can arise from symmetry incompatibility, parameter tuning, or topological interference, and have driven advances in high-Q resonators  \cite{lin2017nonlinear}, low-threshold lasers  \cite{lonvcar2002low,zhang2021halide}, and sensors \cite{tittl2018imaging}.

From a topological perspective, BICs are associated with polarization vortices in momentum space, characterized by integer topological charges \cite{zhang2018observation}. These topological charges are protected against continuous perturbations, offering a powerful framework for understanding and manipulating optical singularities. In addition to BICs, circularly polarized states (C points) are another class of singularities that exhibit half-integer topological charges, enabling the generation and control of chiral optical fields \cite{liu2019circularly,yoda2020generation,chen2021extremize,wang2022realizing}. These polarization singularities form a rich platform for optical vortex manipulation and have attracted considerable attention in the context of chiral light–matter interactions, quantum optics, and structured light \cite{modes2011near}. A particularly intriguing development in this context is the realization of Janus BICs, which exhibit asymmetric topological charges in the upward and downward radiation channels \cite{kang2025janus}. The term Janus describes systems exhibiting asymmetric properties along opposite directions. This concept has inspired new physical phenomena arising from reflection symmetry breaking across diverse platforms, such as electronic materials \cite{montes2022janus,luo2024intrinsic,su20232d,luo2020valley,xu2025ba} and photonic structures \cite{kang2025janus}. This asymmetry provides a new mechanism to break radiation reciprocity while maintaining strong mode confinement, enabling phenomena such as directional emission, spin–momentum locking, and asymmetric light scattering. The first realization of Janus BICs successfully demonstrated that structural symmetry breaking ($\sigma_z$) can effectively induce such radiation asymmetry  \cite{kang2025janus}. This pioneering work not only validates the existence of Janus BICs but also provides a solid foundation for further exploration. 

While structural symmetry breaking provides a viable route to achieving Janus BICs, it naturally raises the question of whether such states can be realized without altering the structural symmetries. Motivated by this, we propose a fundamentally different approach: introducing asymmetry purely at the optical level. This strategy establishes a new design paradigm, where radiation asymmetry and topological charges reversal are achieved through controlled optical asymmetry, rather than structural deformation. The key idea is to construct a geometrically symmetric structure composed of vertically stacked elliptical dielectric cylinders, while introducing optical asymmetry via refractive index contrast between the upper and lower layers. This configuration preserves both in-plane and out-of-plane mirror symmetries at the structural level, yet gives rise to asymmetric radiation behavior due to field asymmetry in the upward and downward directions.

To implement the new design described above, we propose and theoretically demonstrate a symmetry-preserving method to realize Janus BICs on bilayer PCS. In the symmetric configuration, the system simultaneously supports two types of BICs: symmetry-protected BICs located at the Brillouin zone center ($\Gamma$ point), characterized by an antisymmetric field distribution that forbids radiation via parity mismatch, and Friedrich–Wintgen BICs (FW-BICs) at off-$\Gamma$ point, arising from destructive interference between radiative channels. Upon introducing a controlled refractive index detuning between the two layers, the optical symmetry is selectively broken without altering the geometric configuration. As a result, the FW-BICs are destabilized, splitting into a pair of C points with opposite half-integer topological charges ($q = \pm 1/2$), while the symmetry-protected BICs persist. Crucially, the C points associated with the upward and downward radiation shift oppositely in momentum space, leading to a redistribution of topological charges: at the $\Gamma$ point , the upward-radiating channel accumulates a total topological charge of $q = + 1$, while the downward-radiating channel accumulates $q = - 1$. This characteristic of asymmetric polarization vortices and direction-dependent radiation topology constitutes Janus BICs without breaking the structural symmetry. Our proposed strategy enables the robust realization of Janus BICs through a purely optical mechanism without structural deformation, unlocking directional light emission, chiral photonic interfaces, and non-reciprocal photonic elements. These capabilities offer promising opportunities for integrated photonics, optical communication, and quantum information technologies.

\section{Structure design and Computational model}

To explore the formation of Janus BICs, we propose a bilayer PCS model composed of vertically stacked elliptical dielectric cylinders arranged in a square lattice. The lattice constant is set to \emph{a} = 336 nm, and the elliptical cross sections have a diameter \emph{d} = 0.6\emph{a} along the major axis, with an axis ratio of $\beta$ = 1.35. Each cylinder has a height \emph{t} = 240 nm, yielding a total slab thickness of 480 nm. The two layers are directly stacked without an air gap, ensuring strong optical coupling while maintaining both in-plane ($\sigma_x$, $\sigma_y$) and out-of-plane ($\sigma_z$) mirror symmetries in the absence of refractive index detuning. Figures 1(a) and 1(b) present the 3D and top-view schematics, respectively, while Figs. 1(c) and 1(d) illustrate the y-z cross sections for the case of $n_1 = n_2$, and $n_1 \neq n_2$, respectively. Optical asymmetry is introduced by independently tuning the refractive indices $n_1$ and $n_2$ of the lower and upper layers, enabling directional control of the radiation characteristics without breaking geometric symmetries. 

The optical properties of the system are analyzed using numerical finite-element method (FEM) simulations performed in COMSOL Multiphysics. Periodic boundary conditions are applied in the x- and  y-directions to model infinite periodicity, while perfectly matched layers (PMLs) are implemented along the vertical direction to absorb outgoing radiation. The dielectric materials are assumed to be lossless and non-dispersive.  Eigenfrequency analyses are carried out to extract the resonant frequencies, quality factors, and electromagnetic field distributions of the modes. By systematically varying the refractive indices of the upper and lower elliptical cylinders, we explore how optical asymmetry influences the formation of Janus BICs.

In addition, We employ temporal coupled-mode theory (TCMT) to investigate the model characteristics of this system. This complementary approach ensures accurate identification of leaky modes, and their associated polarization structures across the Brillouin zone. We focus on analyzing the modes at different wave vectors \emph{k} and frequencies $\omega$ to identify possible BICs. Particular attention is given to computing the far-field polarization textures and topological charge distributions associated with the leaky modes, which are critical for identifying the emergence of Janus BICs under optical asymmetry. 

	\begin{figure}[h]
		\includegraphics[width=1\linewidth]{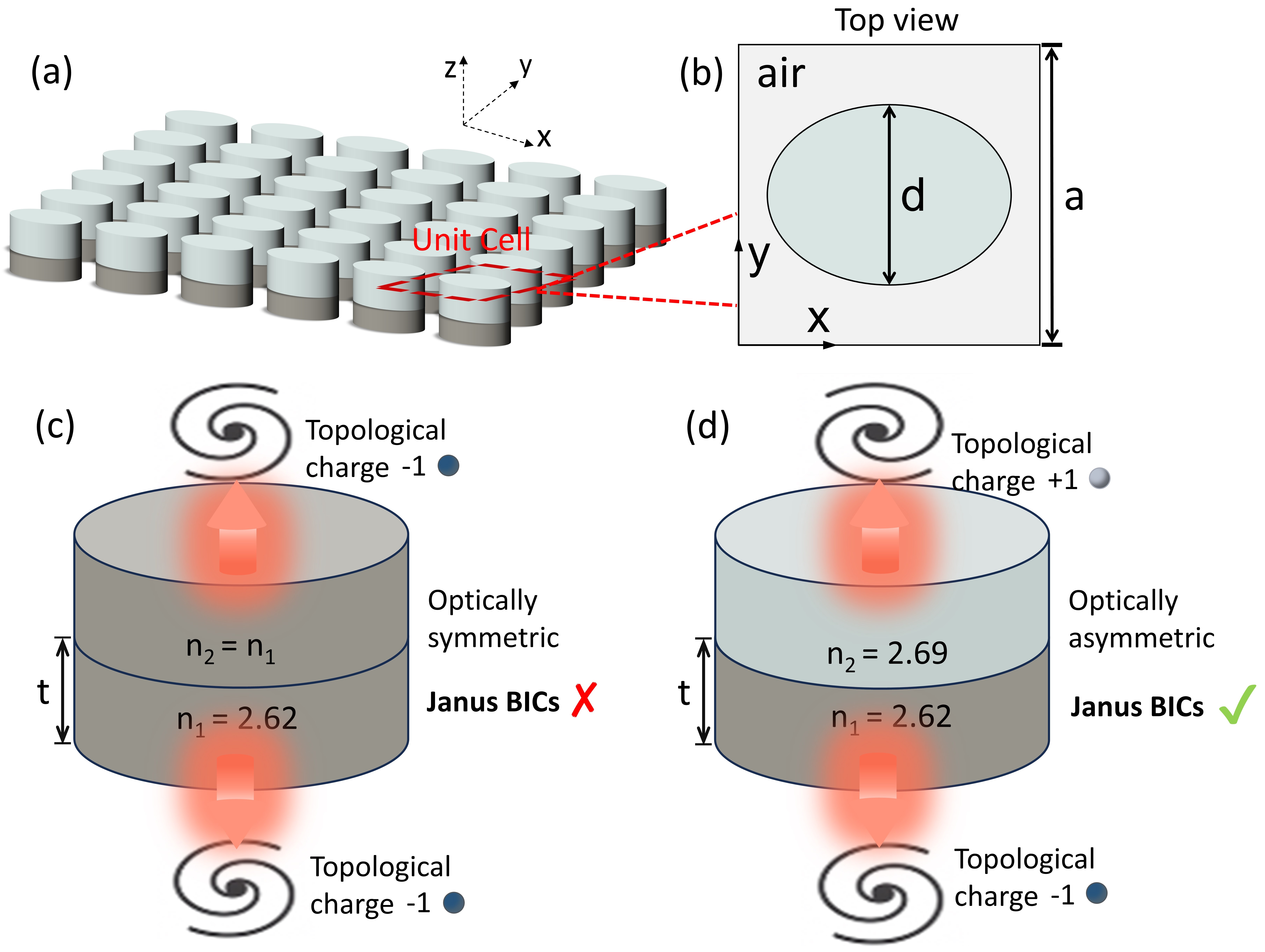}
		\caption{The concept of Janus BICs is explored utilizing highly symmetric PCS. Illustrations of the system in both 3D (a) and top view (b) are presented. The geometric parameters are: \emph{a} = 336 nm, \emph{d} = 0.6\emph{a}, elliptic cylinders with a short- and long-axis ratio of $\beta$ = 1.35, and \emph{t} = 240 nm. The structure consists of bilayer PCS exhibiting $C_{2v}$ in-plane symmetry. (c) A conventional BIC emerges when the refractive index of both layers is set to 2.62. Due to the preserved optical symmetry, the radiation to the upper and lower channels exhibits identical topological charges, preventing the emergence of Janus BICs. (d) By introducing refractive index asymmetry between the two layers while maintaining the in-plane $C_{2v}$ symmetry and out-of-plane geometric symmetry, a pair of optically asymmetric radiation channels emerges. This asymmetry leads to opposite topological charges in the upward and downward far-field radiation, resulting in the formation of Janus BICs.}
	\end{figure}

\begin{figure*}[t]
\includegraphics[width=1\linewidth]{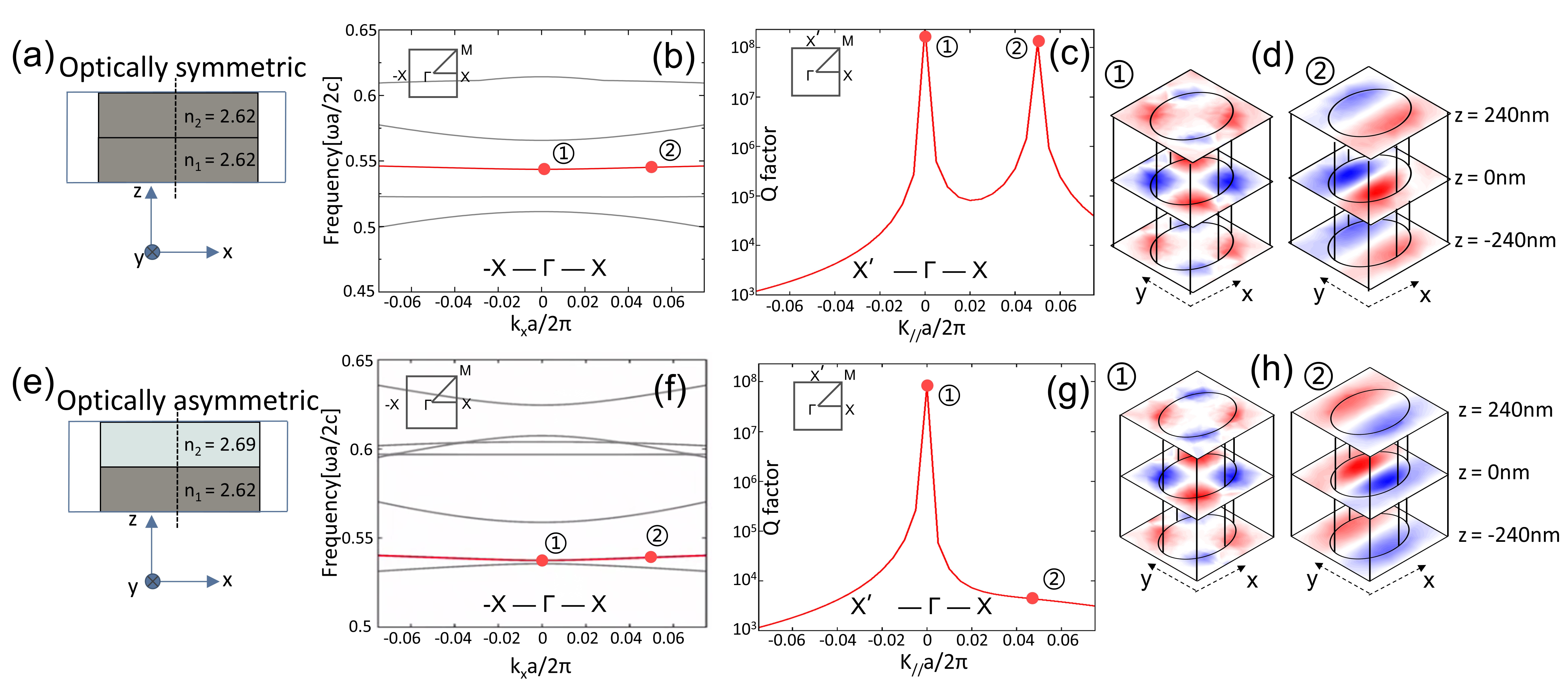}
\caption{ Comparison of optically symmetric and asymmetric configurations in the bilayer PCS. 
(a)–(d) Results for the symmetric case ($n_1$ = $n_2$ = 2.62), where the structure preserves $\sigma_z$ symmetry. (a) Schematic of the symmetric bilayer geometry. (b) Band structure showing two BICs modes labeled \ding{172} and \ding{173}. (c) Corresponding Q factor spectrum indicating that both modes exhibit bound state behavior with ultrahigh Q. (d) Electric field distributions of the two modes reveal mode \ding{172} is a symmetry-protected BIC at the $\Gamma$ point, while mode \ding{173} corresponds to a FW–BICs at an off-$\Gamma$ point. (e)–(h) Results for the optically asymmetric case ($n_1$ = 2.69, $n_2$=2.62), where $\sigma_z$ symmetry of the electromagnetic fields is broken via refractive index detuning. (e) Modified geometry with refractive index contrast between layers. (f) Band structure for the optically asmmtric case. (g) Q factor spectrum indicating that mode \ding{172} (symmetry-protected BICs) remains intact, while mode \ding{173} loses its band character. (h) Electric field profiles confirm that the $\Gamma$ point BICs preserve its antisymmetric mode shape, while the off-$\Gamma$ FW-BICs become leaky due to broken interference conditions. This selective destabilization of FW-BICs under optical asymmetry provides the basis for the emergence of Janus BICs.}
\end{figure*}

\section{Symmetry Analysis and Mode Characterization}

\subsection{Symmetric case: identical refractive indices}
To elucidate the formation mechanisms of BICs in the bilayer PCS, it is essential to analyze the symmetry properties of the structure and the supported optical modes.  In the symmetric configuration, the top and bottom layers have the same refractive index $n_1$ = $n_2$ = 2.62, as shown in Fig. 2(a). Optical modes can thus be classified according to their parity under these mirror operations. As shown in Fig. 2(b), two BICs are supported: one located at the $\Gamma$ point and another at an off-$\Gamma$ point with wave vector $k_x$ = 0.05. These BICs arise near the normalized eigenfrequency $\omega$ $\approx$ 0.547, and are clearly observed in the band structure and the corresponding Q-factor map [Fig. 2(c)]. The field profiles presented in Fig. 2(d) show the $E_z$ component at three representative cross-sections for both BICs. The symmetry-protected BICs \cite{lee2012observation,fan2002analysis,ochiai2001dispersion} at the $\Gamma$ point exhibit an antisymmetric mode profile satisfying $E_{z}(-x,y)$ = $-E_{z}(x,y)$, which forbids coupling to radiation modes due to a parity mismatch. This condition arises from the underlying $C_{2v}$ rotational symmetry in momentum space, which ensures that the eigenmodes of the structure at the $\Gamma$ point can be classified by their even or odd parity with respect to the mirror plane. These BICs are intrinsically robust, as their non-radiative nature stems directly from symmetry mismatch. The off-$\Gamma$ BICs, however, the coupling between modes and free-space radiation becomes more intricate. Modes can radiate into multiple channels characterized by different symmetry properties. Under specific conditions, destructive interference between radiative amplitudes associated with different channels can eliminate radiation leakage, giving rise to (FW-BICs) \cite{ovcharenko2020bound,friedrich1985interfering,gao2023ultrawide}. Such BICs typically occur at off-$\Gamma$ points and are highly sensitive to phase-matching and coupling conditions. In the bilayer PCS studied here, both types of BICs coexist: a symmetry-protected BIC at the $\Gamma$ point arising from parity mismatch, and FW-BICs at finite in-plane wavevectors arising from interference between multiple radiation pathways. This coexistence sets the stage for the emergence of Janus BICs under controlled optical asymmetry, as discussed in the following section.

To elucidate the emergence of Janus BICs in our system, it is essential to first understand the underlying FW-BICs mechanism. We employ TCMT \cite{fan2003temporal,haus1991coupled,suh2004temporal,volya2003non}, a pivotal tool for analyzing interactions between resonant states and traveling waves, which has been extensively applied in the study of weakly coupled multi-ports optical resonator systems. By establishing a coupled dynamical model between localized resonant modes and external radiation channels, this theory provides a rigorous mathematical foundation to model and predict electromagnetic resonance phenomena such as optical BICs. The governing equation for the resonator amplitudes $A$ is given by: \cite{suh2004temporal}:

\begin{equation}
    \frac{\mathrm{d} A}{\mathrm{d} t}=A\left [ i\begin{pmatrix}
 \omega_1  & k\\k
  &\omega_2
\end{pmatrix} -\begin{pmatrix}
  \gamma _1& \gamma _{12} \\
  \gamma _{21}& \gamma _2
\end{pmatrix}\right ] +D^{T}I
\end{equation}

Here, $A$ denotes the amplitude, whose magnitude is determined by the energy stored within the modes. $\omega_{1,2}$ represent the resonant frequencies, and $k$ denotes the near-field coupling rate. The incident wave $I$ from the port couples to the resonator, and the internal resonances decay back into the port with a decay rate matrix $\begin{pmatrix}
  \gamma_1& \gamma_{12} \\
  \gamma_{21}& \gamma_2
\end{pmatrix}$. $\gamma_1$ and $\gamma_2$ correspond to the radiation losses of the two modes, respectively. $\gamma_{12}$ and $\gamma_{21}$, being complex conjugates, characterize the far-field radiation coupling interaction between the two modes.
$D$ represents the coupling coefficient matrix describing the resonance-to-port coupling, which is the transpose of the port-to-resonator coupling coefficient matrix. In this framework, the Hamiltonian matrix of the system is expressed as:\begin{equation}
    i\begin{pmatrix}
 \omega_1  & k\\k
  &\omega_2
\end{pmatrix} -\begin{pmatrix}
  \gamma _1& \gamma _{12} \\
  \gamma _{21}& \gamma _2
\end{pmatrix}=\begin{pmatrix}
 i\omega_1-\gamma_1 & ik-\gamma_{12}\\
 ik-\gamma_{21} &i\omega_2-\gamma_2
\end{pmatrix}
\end{equation}, and solving for its eigenvalues yields the eigenfrequencies and decay rates. Notably, since the frequency and decay of quasi-BICs are described by the complex form $\omega-i\gamma$ (where the central frequency $\omega$ is real), the frequency term in the Eq. 2 is conventionally real-valued, and the effective Hamiltonian matrix is reformulated as: \begin{equation}
     \begin{pmatrix}
 \omega_1  & k\\k
  &\omega_2
\end{pmatrix} +i\begin{pmatrix}
  \gamma _1& \gamma _{12} \\
  \gamma _{21}& \gamma _2
\end{pmatrix}=\begin{pmatrix}
 i\omega_1+\gamma_1 & ik+\gamma_{12}\\
 ik+\gamma_{21} &i\omega_2+\gamma_2
\end{pmatrix}
\end{equation} for analytical tractability.

on the other hand, since the temporal characteristics of the resonance can also be mathematically described by the harmonic dependence $e^{-i\omega t}$ 
, the Hamiltonian matrix of the system in this formalism becomes: \begin{equation}
    \begin{pmatrix}
 \omega_1  & k\\k
  &\omega_2
\end{pmatrix} -i\begin{pmatrix}
  \gamma _1& \sqrt{\gamma_1 \gamma_2}  \\
 \sqrt{\gamma_1 \gamma_2} & \gamma _2
\end{pmatrix}=\begin{pmatrix}
  \omega_1-i\gamma_1&k-i\sqrt[]{\gamma_1 \gamma_2}  \\
 k-i\sqrt[]{\gamma_1 \gamma_2}  & \omega_2-i\gamma_2
\end{pmatrix}
\end{equation} The eigenfrequencies derived from the Hamiltonian matrix in Eq. 4 are given by \begin{equation}
     \begin{split}\omega_{\pm }=&\frac{(\omega_1 + \omega_2)-i(\gamma_1 +\gamma_2)}{2}\pm \\&\frac{\sqrt{\left [   (\omega_1 - \omega_2)-i(\gamma_1 -\gamma_2)\right ]^2 +4(k-i\sqrt{\gamma_1 \gamma_2})^2} }{2}   \end{split}
\end{equation} Only when the condition \begin{equation}
    k(\gamma_1 -\gamma_2)=\sqrt{\gamma_1 \gamma_2}(\omega_1-\omega_2)
\end{equation} is satisfied, the eigenfrequencies represented by Eq. 5 yield purely real values.

\begin{align}
  \omega_+ &= \frac{\omega_1+\omega_2}{2} + \frac{k(\gamma_1+\gamma_2)}{2\sqrt{\gamma_1\gamma_2}} - i(\gamma_1+\gamma_2) \label{eq:omega_plus} \\
  \omega_- &= \frac{\omega_1+\omega_2}{2} - \frac{k(\gamma_1+\gamma_2)}{2\sqrt{\gamma_1\gamma_2}} \label{eq:omega_minus}
\end{align} 

Notably, when $k=0$ or $\gamma_1=\gamma_2$,  BICs occur at $\omega_1=\omega_2$. Such FW-BICs arise purely from mode interference and do not require geometric symmetry, only fine-tuning of material or structural parameters. When these parameters are continuously adjusted to a critical value, the structure supports BICs with a diverging Q factor (Q-infinity). On either side of this critical parameter value, the quality factor decreases, and the BIC transitions into quasi-BICs.

This theoretical framework provides a unified understanding of the emergence of Janus BICs by elucidating the role of FW-BICs via destructive interference between coupled radiation channels. In the symmetric bilayer configuration  (optical asymmetry), FW-BICs arise from complete destructive interference between upward and downward radiation channels. However, when a refractive index contrast is introduced between the two layers, the radiation loss rates $\gamma_1$ and $\gamma_2$ become imbalanced due to the broken $\sigma_z$ symmetry. This imbalance violates the FW-BICs interference condition (Eq. 6), making it no longer possible to fully cancel radiation in both upward and downward directions. Consequently, the ideal FW-BICs evolve into an asymmetric quasi-BICs with finite but directionally distinct radiation. This radiation asymmetry is reflected in the reversal of the topological charges associated with the polarization vortices on the top and bottom surfaces. In this process, the FW-BICs serve as a precursor state that, under perturbation by optical asymmetry, bifurcates into a pair of emission states with contrasting directional properties, marking the emergence of Janus BICs.

\begin{figure*}[t]
    \includegraphics[width=1\linewidth]{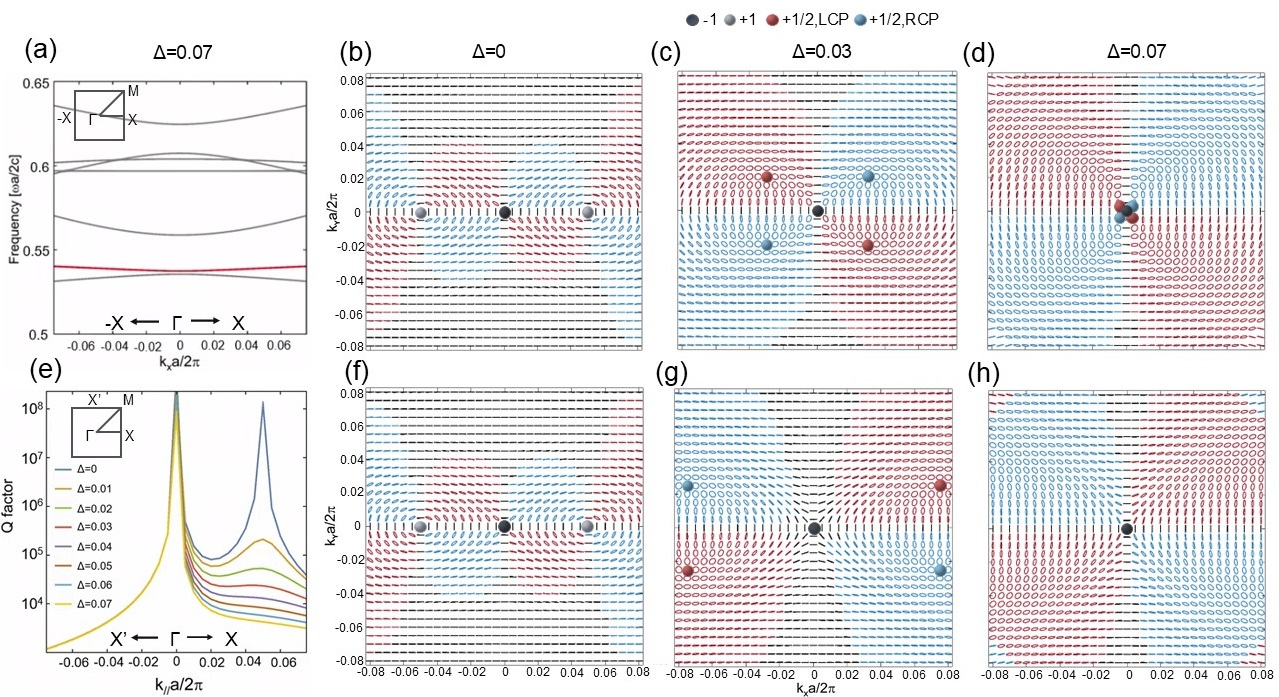}
		\caption{(a) Schematic illustration of the band structure at a refractive index contrast of $\Delta$ = 0.07. (b)-(d) Upward radiating far-field polarization for $\Delta$ = 0, 0.03 and 0.07, respectively. (e) Q factors at different $\Delta$, demonstrating a gradual decrease in the Q factors of the off-$\gamma$ points with increasing $\Delta$. (f)-(h) Downward radiating far-field polarization patterns corresponding to the same $\Delta$ as in (d)-(d). Notably, $\Delta$ = 0.07, the $\Gamma$ point exhibits a topological charge algebraic sum of +1, whereas in (h), this summation yields a value of -1, indicating the emergence of Janus BICs with asymmetric far-field topology.}
  \end{figure*}

\subsection{Asymmetric case: refractive index contrast}

When the refractive indices are detuned($n_1$ $\neq$ $n_2$), the structure  remains geometrically symmetric but becomes optically asymmetric. This optical asymmetry breaks the $\sigma_z$ symmetry of the electromagnetic fields, leading to different far-field coupling conditions for upward and downward radiation. As illustrated in Fig. 2(e), the refractive index contrast between the vertically stacked elliptical cylinders effectively lifts the optical symmetry that supports the off-$\Gamma$ FW-BICs, causing the destructive interference condition to break down. Consequently, the off-$\Gamma$ FW-BICs disappear, resulting in significant radiative leakage and a dramatic reduction in Q factors, as shown in Fig. 2(f) and 2(g). Nevertheless, the symmetry-protected BICs at the $\Gamma$ point persists [Fig. 2(g)], since the in-plane $C_{2v}$ symmetry is preserved. The corresponding field distributions shown in Fig. 2(h) confirm that the antisymmetric mode profile at the $\Gamma$ point is maintained, while the off-$\Gamma$ mode exhibits leakage.
This analysis highlights the distinct physical origins of the two types of BICs: symmetry-protected BICs are robust against certain material perturbations, whereas FW-BICs are sensitive to refractive index asymmetries. Importantly, the ability to control the emergence and stability of BICs via refractive index tuning lays the foundation for engineering Janus BICs with asymmetric radiation properties.


\section{Topological Charge Evolution and Emergence of Janus BICs}

Having established the asymmetric radiation conditions induced by refractive index detuning, we now turn to the evolution of the far-field polarization topology and the emergence of Janus BICs. To understand the radiation properties of photonic modes, especially the emergence of BICs, it is crucial to analyze their behavior in momentum space. While BICs may exhibit spatially localized field pattern in real space, their defining feature is the complete suppression of radiation despite lying within the continuum of radiating states. This non-radiating behavior originates from the far-field interference of outgoing waves in different directions, which is most naturally described in $k$-space. In this context, BICs manifest as polarization vortices in the far-field radiation pattern, where the polarization vector rotates continuously around the singularity. At this vortex center, all radiation channels destructively interfere, leading to a vanishing far-field amplitude and thus forming a non-radiative state. In contrast, radiative modes exhibit uniform or smoothly varying polarization patterns without such topological features. Therefore, the analysis of far-field polarization in momentum space offers a direct and topologically meaningful criterion to identify BICs.

In the symmetric configuration ($n_1$ = $n_2$ = 2.62), mirror symmetry enforces the identical far-field polarization textures in both upward and downward radiation, as illustrated in Figs. 3(b) and 3(f). These topological characteristics are closely associated with the formation of polarization vortices in momentum space. In our system, the BICs located at the $\Gamma$ point carry a topological charge of $q$ = -1, while the off-$\Gamma$ point exhibits a topological charge of $q$ = +1 \cite{ruchi2020phase,qin2023arbitrarily}. These topological charges are equal in magnitude but opposite in sign, forming a polarization vortex pattern. The topological charge $q$ characterizes the number of times the polarization vector winds around a singularity in momentum space and is computed as the winding number of the polarization angle $\phi_{(k)}$ along a closed loop enclosing the singularity 
 \cite{zhen2014topological}:

\begin{equation}
q=\frac{1}{2\pi}\oint_l\,\text dk\cdot \nabla_k\phi (k)   
\end{equation}
\begin{equation}
\phi(k)=arg[C_x(k)+iC_y(k)]  
\end{equation}

Where $\phi(k)$ denotes the angle formed between the major axis of the polarization ellipse and the $k_{x}$ axis in momentum space, characterizing the alignment of the polarization state. Here, $l$ represents a closed loop that is traversed in a counterclockwise direction around the BICs point within the $k_{x}\text{-}k_{y}$ plane.
We use the following formula to represent far-field polarization states \cite{ye2020singular}:
\begin{equation}
     C_x(k)=\frac{1}{\iint_{\text{unit cell}}\,\text {d}x\text {d}y}\iint_{\text{unit cell}}E^*_x\text e^{-\text{i}k_xx-\text ik_yy}\,\text {d}x\text {d}y
\end{equation}
\begin{equation}
   C_y(k)=\frac{1}{\iint_{\text{unit cell}}\,\text {d}x\text {d}y}\iint_{\text{unit cell}}E^*_y\text e^{-\text{i}k_xx-\text ik_yy}\,\text {d}x\text {d}y
\end{equation}

The integration is performed inside a unit cell of an x-y plane slice above (below) the PCS. This formula calculates the Fourier coefficients $C_{x,y}(k)$ of the electric field components $E_x$ and $E_y$ in $k$-space via the Fourier transform, thereby characterizing their spatial frequency content of the electromagnetic field in a periodic structure. 
Specifically, this formula decomposes the electric field in the spatial domain into a superposition of a series of plane waves, each corresponding to a particular wave vector $k(k_x, k_y)$. The coefficient $C_{x,y}(k)$ describes the complex amplitude of the plane wave, quantifying both its strength and phase contribution to the overall field pattern. The exponential term $e^{-\text{i}k_xx-\text ik_yy}$ is the kernel function of the Fourier transform used to convert the spatial coordinates $(x, y)$ into the wave vector space $(k_x, k_y)$, while the normalization factor $\frac{1}{\iint_{\text{unit cell}}\,\text {d}x\text {d}y}$ ensures that the coefficients are independent of the unit cell size. The terms $E_{x}^{\ast }$ and $E_{y}^{\ast }$ in Eq. 11 and Eq. 12 represent the complex conjugate of the electric field component $E_{x}$ and $E_{y}$. This decomposition provides an essential tool for analyzing the radiation characteristics and far-field polarization states of photonic modes, especially in periodic systems where the radiative properties are governed by symmetry and Bloch momentum. 

In our system, to analyze the polarization characteristics of the radiated fields, we utilize the Stokes parameter $S_3$ to determine the handedness of circular polarization with red (blue) denoting left- (right-) handed polarization. This parameter enables the identification of polarization vortices in $k$-space and the calculation of their associated topological charges. The Stokes parameters are defined as follows:
\begin{align}
    &S_{0} = \left | E_{x}  \right | ^{2} + \left | E_{y}  \right | ^{2} \\
    &S_{1} = \left | E_{x}  \right | ^{2} - \left | E_{y}  \right | ^{2} \\
    &S_{2} = 2\,\text{Re}(E_{x}E_{y}^{\ast}) \\
    &S_{3} = 2\,\text{Im}(E_{x}E_{y}^{\ast})
\end{align}
 
The Stokes parameters are four physical quantities that describe the polarization state of light or other electromagnetic waves, including the total light intensity $S_0$, the intensity difference between the horizontal and vertical linear polarization components $S_1$, the intensity difference between the 45° and 135° directional linear polarisation components $S_2$, and the intensity difference between the right-handed and left-handed circular polarization components \cite{kang2023applications,nye1983lines} $S_3$. In our work, the parameter $S_3$ is required to identify circular polarization singularities. In the numerical simulations performed in COMSOL, the electric fields are represented as complex components $C_x$ and $C_y$, leading to the corresponding expression:
\begin{equation}
S_{3} =2Im(C_{x}C_{y}^{\ast }  )  
\end{equation}


When a refractive index detuning $\Delta$ = $n_2$ - $n_1$ is introduced between the bilayers, the out-of-plane optical symmetry is broken. As illustrated in Figs. 3(b)–3(d), increasing $\Delta$ leads to a transformation of the polarization structure in momentum space. Initially, the far-field polarization pattern features a singly charged polarization vortex (with topological charge $q = +1$) located at the off-$\Gamma$ point. This vortex corresponds to a point in $k$-space where the polarization vector winds once around the core as the azimuthal angle $\phi$ varies from $0$ to $2\pi$, indicating a circular distribution of polarization orientations. As $\Delta$ increases, this $q = +1$ vortex becomes unstable under the symmetry-breaking perturbation and eventually splits into two spatially separated C points, each carrying a fractional topological charge of $q = +1/2$. These C points represent phase singularities where the light field is perfectly circularly polarized, and the surrounding polarization field winds with a half-integer strength.

Importantly, the upward- and downward-radiating C points ( those observed in the upper and lower half-spaces) shift in opposite directions in momentum space, as a direct manifestation of the broken mirror symmetry between the two radiation channels. This asymmetric displacement reflects the decoupling of radiation modes along the out-of-plane axis, such that the far-field polarization textures become directionally distinct. In other words, the optical system now supports Janus-type polarization singularities, where the same in-plane mode gives rise to different topological configurations in opposite radiation directions. This asymmetry becomes particularly evident at $\Delta = 0.07$, where the upward-radiating C point approaches the $\Gamma$ point and accumulates a full topological charge of $+1$, while its downward-radiating counterpart exhibits a topological charge of $-1$ [Figs. 3(d) and 3(h)]. Such topological charge imbalance in opposite directions is a hallmark of Janus BICs, characterized by their asymmetric radiation and polarization topology. Notably, this effect arises purely from refractive index modulation, without any structural deformation or explicit geometric symmetry breaking. The Janus BICs thus represent a novel class of topological states: structurally symmetric, but optically asymmetric, characterized by directionally distinct polarization singularities and nonreciprocal radiation topology. This topological asymmetry, while preserving overall geometric symmetry, provides a powerful new degree of freedom for controlling light–matter interactions in photonic systems.

\begin{figure}[h]
    \includegraphics[width=1\linewidth]{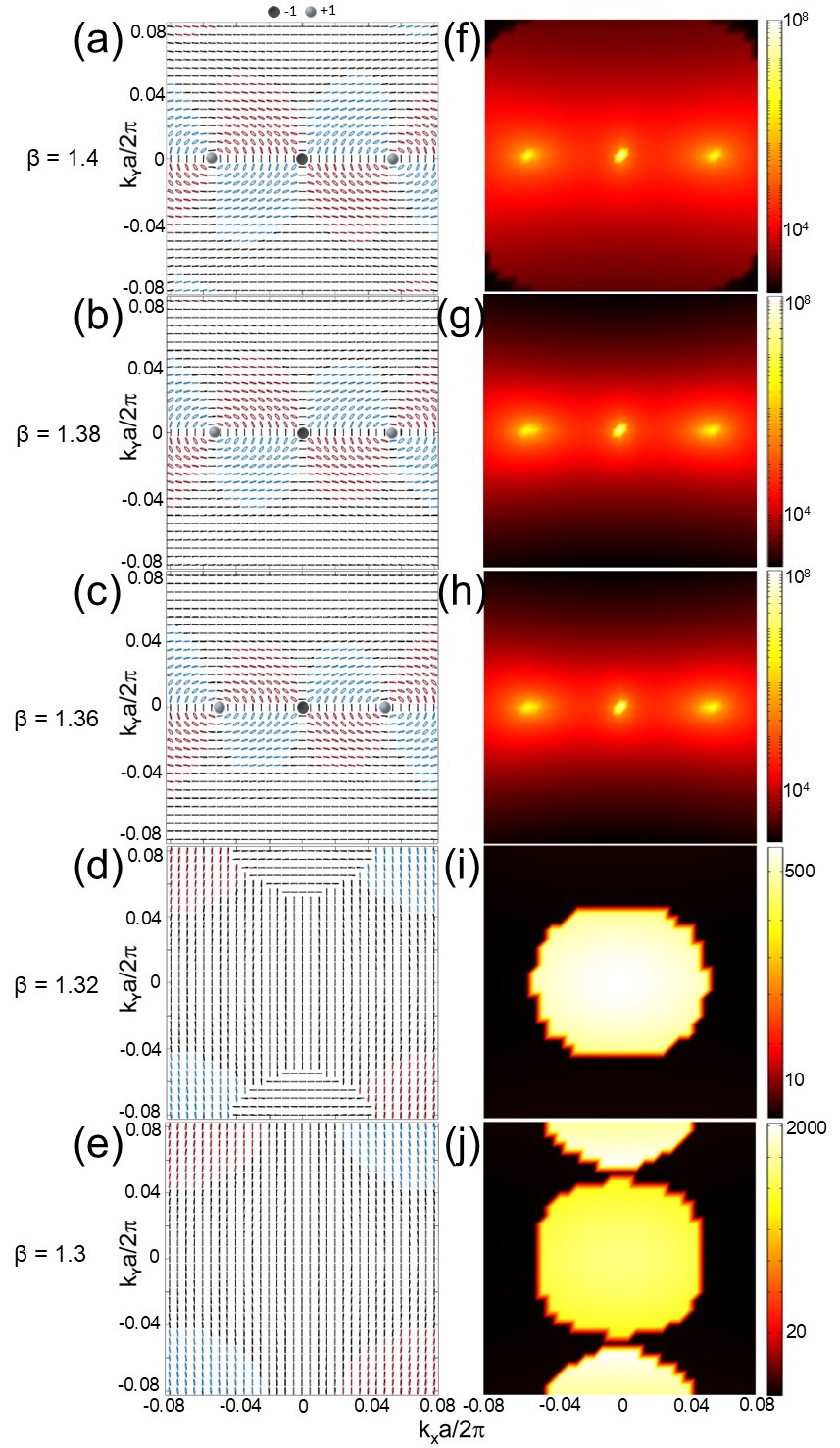}
            \caption{(a)-(e) show the far-field polarisation results for beta equal to 1.4, 1.38, 1.36, 1.32 and 1.3 respectively; at this point there is no breaking of the out-of-plane symmetry, so the far-field polarisation results are the same for the up and down directions. (f)-(j) show the Q factors for beta equal to 1.4, 1.38, 1.36, 1.32 and 1.3 respectively; it can be seen from the figures that the Q factors for beta values in the range 1.36 to 1.4 are very high and vary very little.}
    \end{figure}   

\section{Stability of BICs under Geometric Perturbations}

To further assess the intrinsic stability of the Janus BICs, we investigate the effects of geometric perturbations on their topological and radiative properties. In particular, we focus on variations of the ellipticity parameter $\beta$, which characterizes the ratio between the long and short axes of the elliptical cylinders in the bilayer PCS. We systematically vary  $\beta$ from 1.3 to 1.4, corresponding to realistic parameter fluctuations that may arise during structural design or fabrication processes. For each value of $\beta$, we compute Q factors and far-field polarization distributions of the relevant optical modes. As illustrated in Figs. 4(a)–(c), the far-field polarization patterns for both the upward and downward radiation channels remain identical at the range of $\beta$ from 1.36 to 1.4, indicating that the structural symmetry and the associated singularity topology are preserved. 

\begin{figure*}[t]
    \includegraphics[width=1\linewidth]{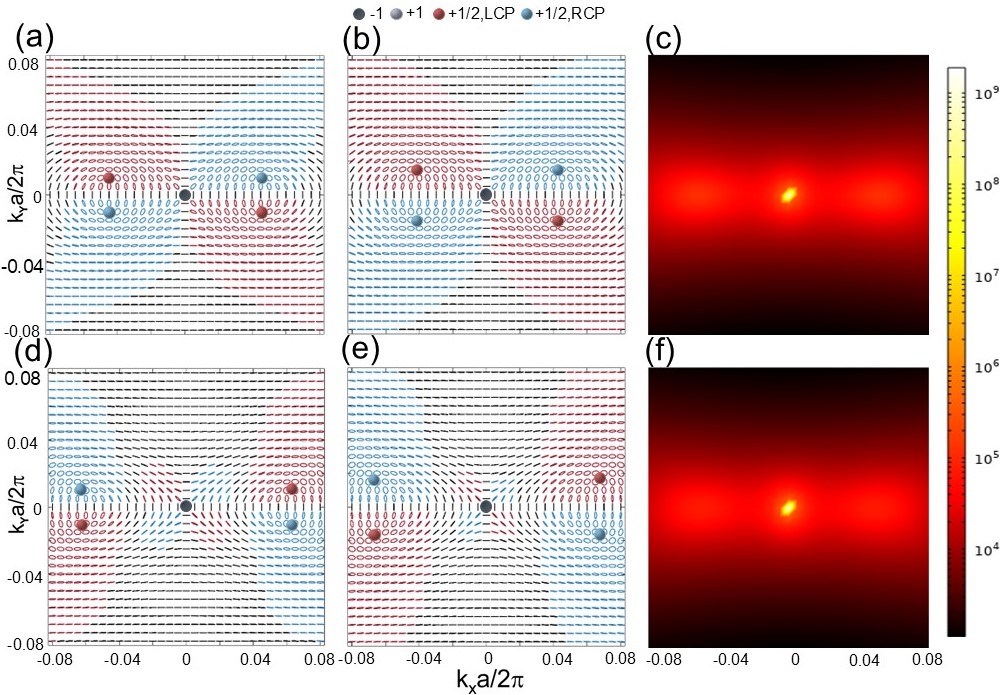}
            \caption{(a) shows the upward-directed far-field polarisation of the first structure (upper elliptical cylinder beta of 1.4 and lower elliptical cylinder beta of 1.35), and (d) shows the down-directed far-field polarisation of the first structure. (b) shows the upward far-field polarisation for the second structure (upper elliptical cylinder beta of 1.42 and lower elliptical cylinder beta of 1.35) and (e) shows the downward far-field polarisation for the second structure. (c) shows the Q factors of the first structure and (f) shows the Q factors of the second structure.}
    \end{figure*}

Correspondingly, the Q factors of the BICs, shown in Figs. 4(f)–(h), remain exceptionally high across this parameter range. This robustness is attributed to the preservation of the $C_{2v}$ in-plane symmetry, which continues to protect the BICs modes from radiative coupling. Crucially, within this range of ellipticity variation, both symmetry-protected BICs at the  $\Gamma$ point and FW-BICs at off- $\Gamma$ points are stably supported. This indicates that the existence of two types of BICs ensures that the necessary conditions for realizing Janus BICs are not restricted to a singular optimized structure, but can be achieved at a range of parameter space. When 
$\beta$ decreases further to 1.32 or 1.3, the polarization patterns are no longer well-defined. As shown in Figs. 4(d)-(e), the far-field polarization field becomes smooth and lacks the singular vortex structure typically associated with topological charges. In such cases, the polarization vectors vary continuously without forming a closed circulation around a singularity, indicating the absence of a well-defined polarization vortex. The corresponding Q factors are also decreased, as shown in Figs. 4(i)-(j). These cases fall outside the regime where BICs can be reliably supported. Consequently, our approach offers considerable flexibility: by appropriately introducing refractive index contrast between the layers in any of these $\beta$ configurations, Janus BICs with asymmetric topological radiation can be reliably achieved. This inherent structural stability enables reliable integration into practical photonic devices, especially in scenarios where environmental fluctuations (e.g., temperature, mechanical stress) or process variations are unavoidable \cite{xing2022capturing}. The ability to sustain high-Q modes under such perturbations underscores the scalability and manufacturability of the Janus BIC platform for real-world applications.

While our primary innovation lies in achieving Janus BICs without structurally breaking the symmetry, relying solely on refractive index detuning to introduce out-of-plane optical asymmetry, it is also instructive to compare this approach with conventional strategies based on geometric asymmetry. Such a comparison not only underscores the advantage of our symmetry-preserving method but also highlights the flexibility and universality of the Janus BICs concept.

To this end, we additionally explore the realization of Janus BICs via structural asymmetry between the upper and lower layers. In this scenario, $\beta$ is varied while keeping the bottom layer fixed at $\beta = 1.35$. In the first configuration, the upper layer has $\beta = 1.4$, and in the second one, it is slightly increased to $\beta = 1.42$. The bottom layer remains unchanged in both cases. The far-field polarization patterns, shown in Figs. 5(a) and 5(d) for the first case, and Figs. 5(b) and 5(e) for the second, reveal the emergence of distinct asymmetric polarization singularities in the upward and downward radiation channels. In particular, the circular polarization singularity in the second structure [Fig. 5(b)] is found to shift closer to the $\Gamma$ point in the upward radiation, while the corresponding singularity in the downward radiation [Fig. 5(e)] moves further away. This spatial separation of polarization vortices with differing topological charges in the two radiation channels reflects the formation of Janus BICs induced purely by geometric means \cite{kang2025janus,zeng2021dynamics}. Despite the broken out-of-plane structural symmetry, the in-plane symmetry remains preserved, ensuring that the modes continue to benefit from symmetry protection. As shown in Figs. 5(c) and 5(f), both structures exhibit ultrahigh Q factors, confirming the bound nature of these states and their minimal radiative loss. 
 This comparison reaffirms that while structural asymmetry remains a valid route to realizing Janus BICs, our method uniquely achieves such functionality while preserving both in-plane and out-of-plane structural symmetry, offering significant advantages in design simplicity and fabrication compatibility.

\section{Conclusion}
In conclusion, we have proposed and theoretically demonstrated a symmetry-preserving strategy for realizing Janus BICs in bilayer PCS. By introducing controlled refractive index detuning between the upper and lower layers, we achieved out-of-plane optical asymmetry without breaking the structural mirror symmetries. This approach enables the coexistence and cooperative evolution of symmetry-protected BICs and FW-BICs, culminating in directionally distinct topological radiation and the formation of Janus BICs. Our results reveal that the upward and downward radiation channels accumulate opposite total topological charges (+1 and -1), a hallmark feature of Janus BICs. Furthermore, we demonstrated that this behavior persists across a range of geometric perturbations, confirming the stability and flexibility of the proposed design. Comparative studies also show that Janus BICs can be realized through structural asymmetry, reinforcing the universality of the concept while highlighting the advantages of our symmetry-preserving method in terms of design simplicity and robustness. These findings establish a versatile framework for engineering topologically nontrivial and directionally controlled photonic states without relying on explicit structural symmetry breaking. The demonstrated resilience and tunability of Janus BICs offer promising opportunities for advancing integrated photonics, directional light sources, nonreciprocal optical components, and chiral light–matter interactions.

\begin{acknowledgements}
H.Y.Meng is supported by the Natural Science Foundation of Hunan Province (2023JJ40612), S.X.X is supported by Changsha Municipal Natural Science Foundation (No. kq2402050), and the Open Project of the State Key Laboratory of Millimeter Waves (Grant No. K202424). 

\end{acknowledgements}

\section*{Author Declarations}

\subsection*{Conflict of Interest}
\noindent The authors declare that there are no conflicts of interest.


\section*{Data Availability}
The data that support the findings of this study are available from the corresponding author upon reasonable request.

\bibliography{Reference}

\end{document}